\documentclass[seceq]{ptptex}
\usepackage{graphicx}
\usepackage{wrapft}
\usepackage[colorlinks=true, pdfstartview=FitV, linkcolor=red, citecolor=blue,urlcolor=cyan]{hyperref}

\title{Fuzzy Bags and Wilson Lines}

\author{Robert D. \textsc{Pisarski}}

\inst{Department of Physics\\
Brookhaven National Laboratory\\
Upton, NY  11973 USA}

\abst{I start with an elementary observation about the pressure
in the deconfined phase of a SU(3) gauge theory without quarks.  
This suggests a ``fuzzy'' bag model for the analogous pressure
in QCD, with dynamical quarks.
I then sketch how the deconfined phase might be described using
an effective theory of Wilson lines.  
To leading order in weak coupling, the effective electric field 
appears in a form familiar from the lattice theory
of Banks and Ukawa.}

\begin{document}
\maketitle

\section{Fuzzy Bags}

The spectacular success of the heavy ion programs at the SPS and
RHIC justifies a careful analysis of the phase transition(s) of QCD
at nonzero temperature.  In this Proceeding I summarize some
recent work of mine,\cite{rdp} hopefully in
a more comprehensible fashion.

Any fundamental understanding of these phase transitions rests upon
the bedrock provided 
by numerical simulations of lattice QCD.  In this section I
begin by looking at old data\cite{three} in a new way.
This was mentioned in a footnote, Ref. 37 of Ref. 1.
While the data is for a pure gauge theory, 
it shows that except very near $T_c$, the critical temperature, there
is a exceedingly simple form for the pressure.
This immediately suggests an ansatz applicable to QCD,
and which might be of use for phenomenology.
\begin{figure}
  \begin{center}
       \includegraphics[width=\halftext]{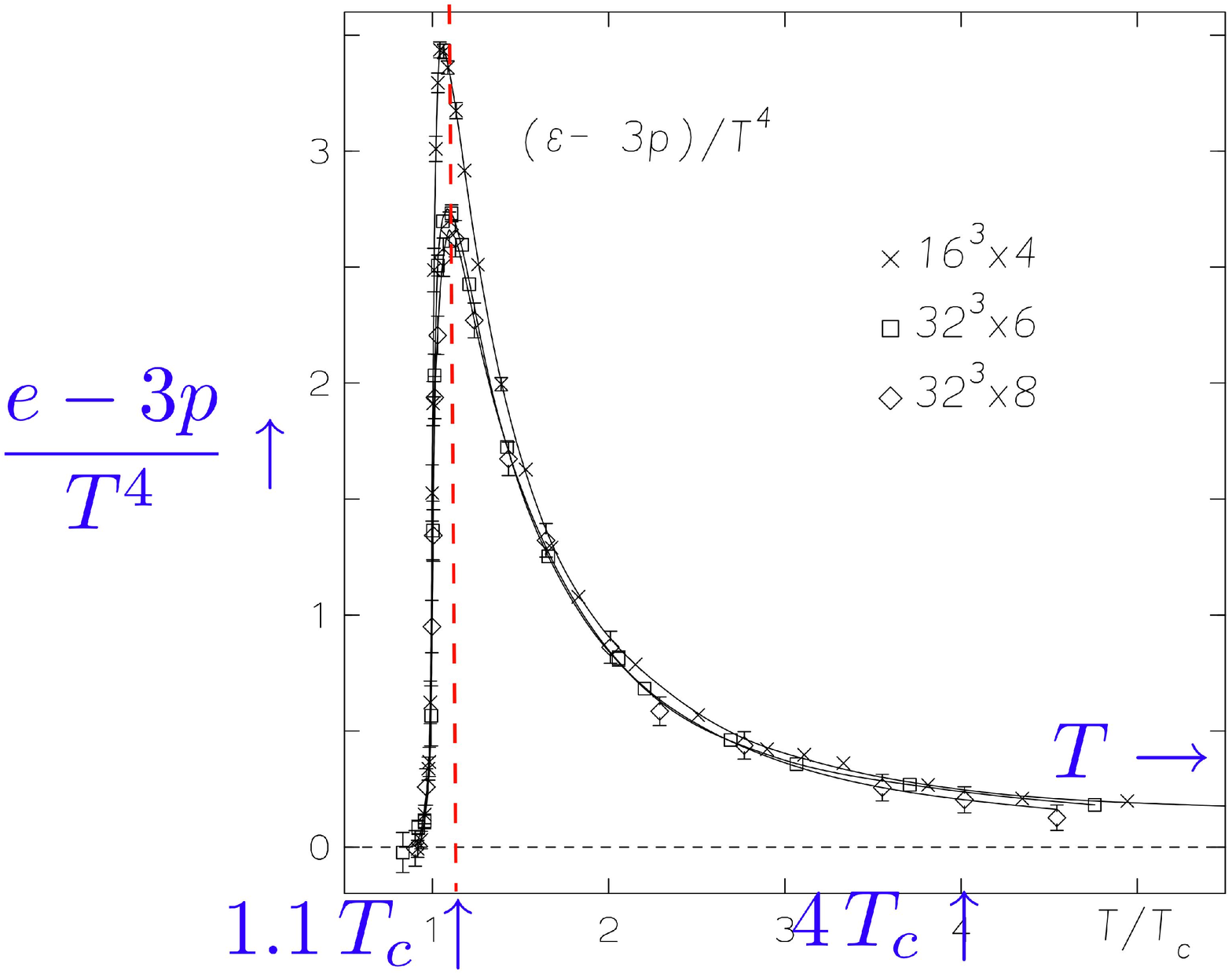}
       \includegraphics[width=\halftext]{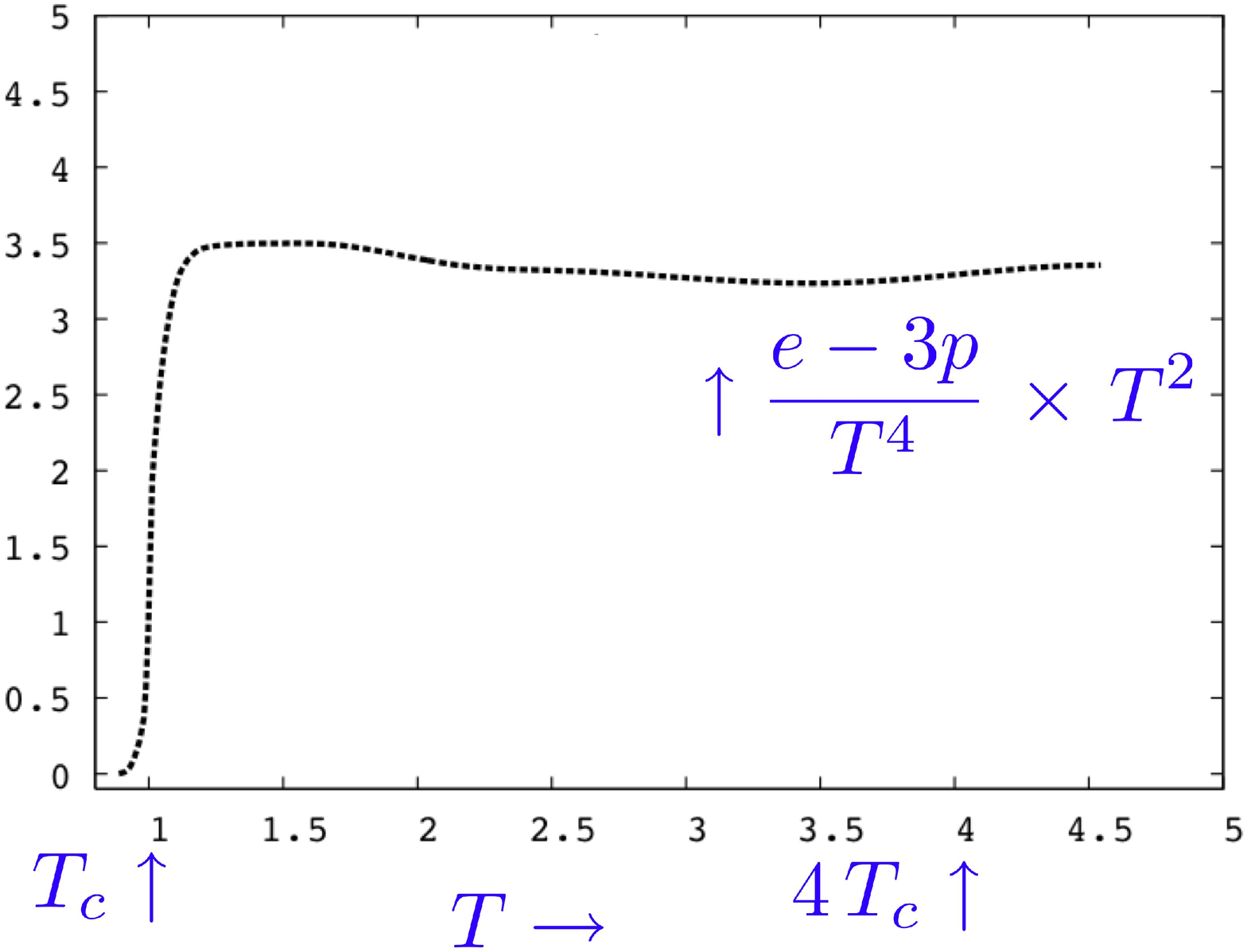}
    \caption[]{Lattice results\cite{three} for the pure SU(3) gauge theory:
to the left, $(e-3p)/T^4$; to the right, the same quantity times $T^2$.
}
    \label{fig:pressure}
  \end{center}
\end{figure}

Ten years ago, a group at Bielefeld\cite{three} 
computed the thermodynamic properties 
of a SU(3) gauge theory, close to the continuum limit.
In equilibrium, while all 
thermodynamic quantities follow from the pressure, 
it is convenient to plot what is usually called the ``interaction
measure'', $(e-3p)/T^4$, where
$e$ is the energy density, $p$ is the pressure, and $T$ the temperature.
This is plotted in the left panel in Fig. 1.
The interaction measure is the trace of the energy momentum tensor, divided by
$T^4$, and so vanishes if the theory is conformally symmetric.
A pedestrian way of seeing this is to note that
$e - 3p = T(dp/dT) - 4 p$; if the theory is conformally invariant,
the pressure is just a pure number times
$T^4$, for which $e-3p$ vanishes.

Thus the interaction measure is a dimensionless
number which quantifies the deviation from conformal symmetry.
Because the conformal anomaly is proportional to the $\beta$-function,
this also measures the deviation from ideality.
As can be seen from the left panel, the
interaction measure is very small below $T_c$, 
rises steeply around $T_c$, with a sharp peak at
$T_{\rm max} \approx 1.1 \, T_c$.  Above $T_{\rm max}$ it trails
off relatively slowly.  By $T_{\rm pert} \approx 4.0 \, T_c$, 
its value is, within a factor of two,
equal to that expected from perturbation theory,\cite{rdp}
where the interaction measure begins as $\sim \alpha_s^2$.

My concern is with the fall off between $T_{\rm max}$
and $T_{\rm pert}$.  In the right panel of Fig. 1, I take $(e-3p)/T^4$,
and simply multiply times $T^2$.  As can be seen, 
$(e-3p)/T^2$ is essentially constant.  
For the temperatures shown, then, the pressure is a sum of just {\it two}
pieces: an ideal term, $\sim T^4$, and 
a new, non-ideal term, $\sim T^2$.  

If one ignores the overall
normalization, the relative normalization of the two terms can
be computed without further ado.
In an asymptotically free theory, at high temperature the pressure
approaches that of an ideal gas, so at any temperature it
is natural to compare the pressure to that of the appropriate ideal gas.
In the pure glue theory, $T_c \sim 270$~MeV is much smaller 
than the lightest glueball mass, 
$\sim 1.5$~GeV, and so, relative to the ideal gas,
the confined pressure is very small.\cite{three,two,four} 
Thus the pressure (nearly) vanishes at $T_c$.  This gives
\begin{equation}
p_{\rm pure \;glue}(T) 
\approx f_{\rm pert} \left( T^4 - T_c^2 \, T^2 \right) \;\;\; , \;\;\;
 1.1 \,T_c \leq T \leq 4.0 \, T_c \; ,
\label{glue}
\end{equation}
where $f_{\rm pert}$ is a constant.

What of other numbers of colors, $N_c$?  The deconfining phase transition
is of second order\cite{two} for $N_c =2$, and
of first order\cite{three,four} when $N_c \geq 3$.  As $N_c$ increases,
the transition becomes more strongly first order,\cite{four}
with a latent heat $\sim N_c^2$.
For three colors,\cite{three} 
the transition is weakly first order.

The change in the order of the transition with $N_c$
affects the interaction measure, but not dramatically so.\cite{four}
Since the pressure is continuous at $T_c$, and as the confined
phase has negligible pressure,
then whatever the order, the pressure is almost zero at $T_c$,
$p(T_c) \approx 0$.  In contrast,
the energy is sensitive to the order: while $e(T_c^-) \approx 0$,
just above the transition the energy 
vanishes for a second order transition, and
is nonzero for a first order transition.

For a first order transition, then, at $T_c$ the interaction
measure $\approx e(T_c^+)/T_c^4$.  For the strongly first order
transition with four colors,\cite{four} it is not surprising to find
that the maximum in the
interaction measure is at $T_c = T_{\rm max}$, and that it
falls off after that.

For a second order transition, the energy is continuous at $T_c$,
so $e(T_c^+) \approx 0$, and the interaction measure is nearly zero
at $T_c$.  For the deconfining transition with two colors,\cite{two}  
the interaction measure increases from near zero at $T_c$, has
a sharp maximum at $T_{\rm max} \approx 1.15 \, T_c$, 
and then falls off after that.
Thus for two colors,\cite{two} the peak
in the interaction measure can be viewed
as a remnant of that for an infinite number of colors.\cite{four}
Since the transition is nearly second order
for three colors,\cite{three} $e(T_c^+)$ is
small, and the interaction measure
looks like that of two colors; the peak in the interaction
measure moves closer to $T_c$, to $T_{\rm max} \approx 1.1 \, T_c$.

The formula in (\ref{glue}) only applies above the maximum
in the interaction measure, 
so we should only compare different $N_c$ at $T > T_{\rm max}$.
For $N_c = 4$, this is for
all $T \geq T_{\rm max} = T_c$.  This appears to be supported
by lattice simulations.\cite{four}
To be fair, the data for $N_c = 8$ does not, but perhaps
lattice discretization errors are larger there.\cite{four}
For two colors, one compares for $T > T_{\rm max} = 1.15 \, T_c $:
the data of Ref. 3 appears to violate
(\ref{glue}) by a large amount, $\sim 50\%$.
However, it is not clear that these simulations are 
close to the continuum limit, and so 
new simulations would be most welcome.

Before describing how one might extend (\ref{glue}) to
other temperatures,
to $T < T_{\rm max}$ and $T> T_{\rm pert}$, I skip ahead directly to
the case of dynamical quarks and three colors.
For three flavors or less,\cite{newthree} while
the ``transition'' often becomes a crossover,
an approximate ``$T_c$'' can still be defined.
Whatever the order, though, the interaction measure behaves similarly.
Relative to an ideal gas of quarks and gluons,
there is a small but significant pressure below $T_c$,
and a maximum in the interaction measure at a
temperature $T_{\rm max}$, which is above $T_c$.
The surprise about (\ref{glue}) is that 
the leading correction to the ideal gas term is $\sim T^2$, and not
$\sim T^3$.  I speculate that this is generic: that for
$T_{\rm max} < T < T_{pert}$, 
the pressure is a series in powers of
$1/T^2$ times the ideal $T^4$ term.
I call this a ``fuzzy'' bag model for the pressure:
\begin{equation}
p_{\rm QCD}(T) 
\approx f_{\rm pert} \; T^4 - B_{\rm fuzzy} \; T^2 
- B_{\rm MIT} + \ldots \;\;\; , \;\;\;
T_{\rm max} \leq T \leq  T_{\rm pert} \,  \; 
\label{fuzzy}
\end{equation}
The upper bound, $T_{\rm pert}$, denotes the point at which
perturbative contributions to the interaction measure are of the same
order as that found from the lattice; it
is a few times $T_c$, something like $\approx 4 T_c$.

In (\ref{glue}), $f_{\rm pert}$ is dimensionless, $B_{\rm MIT}$
is the usual MIT bag constant,\cite{MIT} with dimensions of $({\rm mass})^4$,
and $B_{\rm fuzzy}$ is a fuzzy bag constant, with dimensions
of $({\rm mass})^2$.  For the pure glue theory, from (\ref{glue}) 
$B_{\rm fuzzy} = f_{\rm pert} T_c^2$, and $B_{\rm MIT} \ll B_{\rm fuzzy}$.

Recent lattice simulations\cite{newthree} appear to support
(\ref{fuzzy}).  Note that with a fuzzy bag constant, the sign
of the MIT bag constant is not guaranteed, 
but these simulations find that $B_{\rm MIT}$ is positive,\cite{newthree}
as in the original MIT bag model.\cite{MIT}  

With a little work, it should be possible to generalize 
(\ref{fuzzy}) to the entire deconfined phase. Above $T_{\rm pert}$,
we can ignore the non-ideal terms, and include only $f_{\rm pert}$,
now considered as a function of $T$.
This is given by resummations of weak 
coupling perturbation theory at nonzero 
temperature.\cite{resum,helsinki}  While all resummations fail
at temperatures below $T_{\rm pert}$, in the present
view this is just because of the {\it non}-ideal terms in the
pressure.  This suggests that for all $T \geq T_c$,
perturbative resummations contribute
only to $f_{\rm pert}(T)$.  Analysis shows that
$f_{\rm pert}(T_{\rm pert})$ is about
$90\%$ of the ideal gas value.\cite{resum,helsinki} 
If applied just to $f_{\rm pert}(T)$, it seems very possible that
perturbative resummation might work all of the way down to the
critical temperature.\cite{resum}
With present day techniques, lattice simulations could
test this in a precise manner in the pure glue theory. 

This suggests a heuristic analogy,
to the operator product expansion for two gauge invariant operators
at short distances.
Free field theory dominates as the distance
$x\rightarrow 0$, like some power of $x$.  
Perturbative corrections enter with the same
power, times a series in $1/\log(x)$, {\it etc.}
Non-perturbative effects involve the expectation
values of new operators, multiplied by the appropriate
powers of $x^2$.  For the pressure, perturbative terms correct
the ideal $T^4$ term, as a series in $1/\log(T)$, {\it etc.},
while non-perturbative effects generate non-ideal
terms, as a series in $\sim 1/T^2$ times the ideal term.

What about below $T_{\rm max}$?  
I assume that a hadron
resonance gas provides a reasonable approximation not just about
$T=0$, but all of the way to $T_c$.  This leaves
$T_c \leq T \leq T_{\rm max}$.  
For three colors in the pure glue theory, this is a nearly critical
regime, dominated by a light excitation for the triplet Polyakov loop.
Maybe even with dynamical quarks, this region is dominated by 
a light triplet loop and its interactions with quarks.\cite{loop}

Why bother?  For a pure glue theory, simulations
relatively close to the continuum limit have been possible for some
time.\cite{three}  
With dynamical quarks, though, and in particular for the light quarks
present in QCD, present day
simulations are not close to the continuum
limit.  Thus (\ref{fuzzy}) could be used by approximate models.
For example, most
hydrodynamic models\cite{hydro} use a MIT bag model for
the pressure.  This approaches
ideality much faster than a fuzzy bag
model.  Since hydrodynamics only requires the energy as a function of
pressure, perhaps non-ideal corrections don't
really matter that much, but this should be demonstrated by explicit
computation.

Terming (\ref{fuzzy}) a ``fuzzy'' bag is not sheer whimsy.
In the MIT bag model the surface of the bag is infinitely thin
and fixed.\cite{MIT}
While the interface surely has nonzero width, it is difficult to know
how to model this.  For instance, if the
surface of the bag were thin and floppy, then there would be 
many light excitations,
in which the surface of the
bag flops around, and the quarks remain essentially
fixed.\cite{jaffe}
There are no signs of such a multiplicity
of states in the hadronic mass spectrum.

The essential moral of the non-ideal terms in the pressure, 
(\ref{glue}) and (\ref{fuzzy}), is that the transition from a confined,
to a nearly perturbative phase, is not abrupt, as in a MIT bag model,
but gradual.  This suggests that the boundary of the bag isn't
thin, but thick.  A thick bag is unlikely to flop around, since
the entire width needs to participate.  Thus light surface modes
shouldn't be a problem.

Having said this, it is not clear how to develop a more realistic bag,
with a thick boundary.  The thickness probably doesn't affect the mass
spectrum of ordinary hadrons greatly.  The analogy
is still suggestive.  In some loose sense
nonzero temperature probes inverse distances in the QCD vacuum, $T \sim 1/R$:
the perturbative vacuum emerges
as $T\rightarrow \infty$, or $R\rightarrow 0$; the confined vacuum,
for $T\rightarrow 0$, or $R\rightarrow \infty$.  The width of the
bag emerges over distances 
$R \approx 1/T_{\rm pert} \rightarrow 1/T_{c}$; in 
physical units, for $\approx 1/4 \; {\rm fm} \, \rightarrow 1 \; {\rm fm}$.
This is the right scale to probe the transition from going inside, to outside,
the bag.

\section{Wilson lines and their electric field}

The above begs the question: what is the origin of the non-ideal terms
in the pressure of the deconfined phase?  I next turn to a possible
explanation in terms of Wilson lines.\cite{rdp}  
I omit all details, and many references, to concentrate
on a broad and qualitative description.

In resummations of perturbation theory,\cite{resum,helsinki} 
the gluon degrees of freedom remain
the time-like, $A_0$, and space-like, $A_i$, 
components of the vector potential.  To proceed further, 
consider a straight, thermal Wilson line:
\vspace{-.05in}
\begin{equation}
L(x) = {\textit {\large P}} \;
{\textit {\Large e}}^{\; \displaystyle i  g
\displaystyle \int^{1/T}_0 \displaystyle A_0(x,\tau) \; d\tau } \; .
\label{def_wilson_line}
\end{equation}
$\tau:0\rightarrow 1/T$ is the imaginary time, $x$ is the spatial coordinate,
and $g$ the gauge coupling constant.
If $A_0$ is in the fundamental representation, 
$L$ transforms as an adjoint matrix: under strictly periodic gauge
transformations, ${\cal U}(x,0) = {\cal U}(x,1/T) = {\cal U}(x)$,
$L(x) \rightarrow {\cal U}^\dagger(x) L(x) \, {\cal U}(x)$.

The trace of the Wilson line is the Polyakov loop, and is invariant
under local gauge transformations.  
The motivation for considering an effective theory of the Wilson line
rests upon measurements of the renormalized Polyakov loop.  In
a perturbative regime, fluctuations in $A_0$ should be small,
and so suitably normalized, this expectation value should be near one.
In contrast, if fluctuations in $A_0$ are large, the expectation of
the Polyakov loop is not near one.  Lattice simulations in
a SU(3) gauge theory, with or without quarks, show
the (renormalized, triplet) loop is near one for $T > T_{\rm pert}$,
and less than one for $T_c < T_{\rm pert}$.

One's first guess might be that the theory is driven into a regime of
strong coupling.  Consider, however, 
the ``Helsinki'' program of resummation.\cite{helsinki} 
Originally proposed by Braaten and Nieto, computations
to four loop order were done by
Kajantie, Laine, Rummukainen, and Schr\"oder.
The final steps are being completed by
Laine, Schr\"oder {\it et al.}\cite{helsinki}  They find that
that the effective coupling runs with a mass scale $\sim 2 \pi T$,
so that even down to $T_c \sim 175$~MeV, the
QCD coupling is $\alpha_s^{\rm eff}(1.6 \,{\rm GeV}) \sim 0.28$.
This value is {\it not} that large.\cite{helsinki}

This suggests the perturbative construction of an
an effective theory in three dimensions,
valid over distances $> 1/T$.  Since
the renormalized Polyakov loop is not near unity, 
$A_0$ is replaced by Wilson lines, coupled as always to the $A_i$.

When I first suggested this there were several problems.\cite{loop}
The first is how one to match the effective theory, at large
$A_0$, to quantities which are computable perturbatively.
In the Helsinki program, this is done by computing the positions of
poles in propagators, {\it etc.}  This is fine at small $A_0$,
but doesn't probe large $A_0$.

Interfaces\cite{altes} can be used to probe large $A_0$.
These are most familiar in a pure gauge theory, such as SU(N).
A SU(N) gauge group has a global center symmetry of Z(N),
so that in the deconfined phase, there are $N$ degenerate vacua:
the usual perturbative vacua, $L = 1_N$, and Z(N) transforms
thereof, such as $L = \exp(2 \pi i/N) \,1_N$.

An interface interpolates between these degenerate vacua.  One
takes a box which is long in one spatial direction, say $z: 0 \rightarrow z_f$.
At one end of the box, one takes one vacua, 
$L(0) = 1_N$; at the other end, an inequivalent Z(N) state, 
$L(z_f) = \exp(2 \pi i/N) \, 1_N$.  These boundary conditions force
the formation of an interface along
$z$, which tunnels between the degenerate vacua.
While the ends of the box represent perturbative vacua, in between one probes
large $A_0 \sim T/g$, as illustrated by (\ref{A0transf}).
The amplitude for tunneling can be computed
semiclassically, and gives an interface tension 
$\sim T^2/\sqrt{g^2}$.  
The width of the interface is 
proportional to the inverse Debye mass, 
$\sim 1/(\sqrt{g^2} T)$, so that a derivative
expansion can be used. To date, computations have
been carried out to $\sim (\sqrt{g^2})^{3/2}$ times 
the result at leading order by Giovannangeli and Korthals Altes.\cite{altes}  

Such interfaces appear to be special to theories with a center symmetry,
and so useless for QCD, where it is violated by the
presence of dynamical quarks.  In this case, however,
there are other interfaces which can be used.
Consider the gauge transformation
\begin{equation}
{\cal U}_c(\tau) =
\textit{{\Large e}}^{\,
\textstyle 2 \pi i \, \tau T \, t_N/N} \;\;\; , \;\;\;
t_N =
\left(
\begin{array}{cc}
1_{N-1} & 0      \\
0             & -(N-1) \\
\end{array}
\right) \; .
\label{gauge_transf}
\end{equation}
This is only periodic up to a Z(N) rotation,
${\cal U}_c(1/T) = \exp(2 \pi i/N) \, {\cal U}_c(0)$, which
is allowed in the absence of quarks.
If one acts with ${\cal U}_c$ on one end of the box, but not the other,
then a Z(N) interface forms, because this
isn't a pure gauge transformation
in between the two ends.

Now instead of (\ref{gauge_transf}), 
consider the $N^{th}$ power thereof, ${\cal U}_c^N$.
This is a strictly periodic gauge transformation, and thus
is allowed, independent of whatever matter fields may be present.
If we act with ${\cal U}_c^N$ on one end of a box, but not
the other, what I term
a U(1) interface is generated: while $\langle L \rangle = 1_N$
at both ends, in between $A_0$ winds around in a topologically
nontrival fashion.

To compute the properties of
an interface, one needs the effective potential for
constant $A_0$, which is first generated at one loop order.
For a effective theory of Wilson lines, it is obvious to turn 
an effective potential of $A_0$ into one for $L$.  What stymied me\cite{loop}
is what one does at {\it zeroth} order: how does one write
the effective electric field in terms of Wilson lines?

The crucial clue was provided by what appeared to
be an abstruse computation.
At one and two loop order in a SU(N) gauge theory, explicit
calculation shows that
the effective potential, computed in the presence of a 
large, background field for constant $A_0$ --- and $A_i = 0$ ---
is invariant under the Z(N) center symmetry.\cite{altes}
This is unremarkable: there is no anomaly to prevent
the quantum theory from respecting the center symmetry of the 
classical theory.  Diakonov and Oswald\cite{diakonov} then
computed in the presence of background
fields in which both $A_0$ {\it and} $A_i$ were nonzero, 
allowing $A_0$ to be large.  Assuming that the effective
electric field is $D_i A_0$, they found that the center
symmetry appeared to be violated at one loop order.
An error in computation seems unlikely, given that
their results agree with those of Megias, Ruiz, and Salcedo,\cite{megias}
who computed in the limit of small $A_0$ and $A_i$.

I suggest that the problem lies
in assuming that the effective electric field
is $D_i A_0$, and arises even at tree level.
Under the large gauge transformation of (\ref{gauge_transf}),
\begin{equation}
A_0^{diag} \rightarrow A^{diag}_0 + \frac{2 \pi T }{g N} \; t_N
\;\;\; , \;\;\; A_i \rightarrow {\cal U}_c^\dagger \; A_i \; {\cal U}_c \; .
\label{A0transf}
\end{equation}
Hence diagonal elements of $A_0$ are shifted by a large but constant
amount, $\sim T t_N/g$, while off-diagonal elements of $A_0$ and
$A_i$ undergo time dependent rotations. 

In four dimensions, the original electric field is 
$D_i A_0 - \partial_0 A_i$.
The first term, $D_i A_0 = \partial_i A_0 - i g [A_i,A_0]$,
changes if $[A_i,t_N] \neq 0$, which always happens
for some (off-diagonal) components of $A_i$.  
It is easy to see, however, that the
time dependent rotation of $A_i$ in the second term, $-\partial_0 A_i$,
generates a similar commutator, $-[A_i,t_N]$, which exactly cancels
against the first term.  This just reflects the fact that the
original electric field transforms homogeneously under
arbitrary gauge transformations.

In the effective theory, the simplest guess for the effective electric
field is to drop time derivatives, and take it to be $D_i A_0$.
While valid at small $A_0$,\cite{resum,helsinki} this can't be right
at large $A_0$.  The above argument shows that
$D_i A_0$ is not invariant
under the large gauge transformations, ${\cal U}_c$, which enforce
the center symmetry.\cite{diakonov}
Similarly, if we take ${\cal U}_c^N$, 
the effective theory is not even invariant under large, but strictly
periodic, gauge transformations.

The resolution is to construct the effective electric field from
Wilson lines.  These transform like a phase under ${\cal U}_c$,
and are invariant under ${\cal U}_c^N$.
To leading order in weak coupling, the effective electric field is
\begin{equation}
E^{\rm eff}_i(x) = \frac{T}{ig} \; L^\dagger(x) \, D_i \, L(x) \; .
\label{electric}
\end{equation}
This is shown by demonstrating that the interface tensions,
either Z(N) or U(1), agree between the effective and original theories.
Eq. (\ref{electric}) does change beyond leading order, and is
multiplied by other gauge invariant terms, 
such as $|{\rm tr} \, L|^2$, $|{\rm tr} \, L^2|^2$, {\it etc.};
with coefficients which begin at $\sim g^2$.

It is well known that the mapping between the fields in an effective theory,
and the original theory, is indirect.  Usually, however, one only
sees this at next to leading order in some expansion,
so the differences are small.  In the present instance, because
one is constructing an effective theory for large $A_0$, it arises
even at zeroth order.  
The problems found before\cite{diakonov} are presumably solved
by matching to an effective theory constructed from the effective
electric field, and not functions of $D_i A_0$.

To leading order, the effective Lagrangian is
that of a gauged, principal chiral field:
\begin{equation}
{\cal L}^{\it eff}_{\it classical}(A_i,L) =
\; \frac{1}{2} \; {\rm tr}\; G_{i j}^2 + \frac{T^2}{g^2}
\; {\rm tr}\, |L^\dagger D_i L|^2 \; .
\label{classical}
\end{equation}
This is nonrenormalizable, but this shouldn't be a problem, since
the effective theory is only valid over distances $> 1/T$.
A related linear model, constructed to be
renormalizable, has been analyzed
by Vuorinen and Yaffe.\cite{vuorinen}

On the lattice, the Lagrangian of (\ref{classical}) was first written
down by Banks and Ukawa,\cite{banks} as the simplest kinetic
term for an adjoint field, $L$.
It is not obvious that (\ref{classical}) applies
in the continuum, even at leading order in weak coupling.
It should be possible to construct effective Lagrangians to the
same order as interfaces, in the original
theory, have been computed.\cite{altes}

One can show that the four dimensional instanton number 
equals the winding number of the Wilson line.\cite{rdp}
Dai and Nair\cite{nair} showed that non-abelian hydrodynamics has
color Skyrmions.  This suggests that the effective model might have
electric Skyrmions:\cite{nair} solutions stabilized by a nonzero winding
number for $L$, and yet which are not instantons, since they only carry
electric, and not magnetic, fields.

How does the deconfining transition arise?
For the potential for $L$ computed perturbatively, order by
order the deconfined vacuum is
always proportional to the unit matrix, $\langle L \rangle \sim 1_N$.
This could be possible all of the way to $T_c$ --- but then one would
not expect non-ideal terms in the pressure.  To destabilize the perturbative
vacuum, it is necessary to add, by hand, a non-perturbative
mass term for the Wilson line.  The simplest example is
$\sim T^2 \, B_f \,|{\rm tr}\,L  |^2$.  The notation $B_f$, and the $T^2$,
is motivated by the previous section; adding such a mass term is
standard in a Landau-Ginzburg type of analysis.

With such a term, $\langle L \rangle \neq 1$.  
In Ref. 1, I argued that at infinite $N_c$,
the confined vacuum is characterized by complete eigenvalue repulsion;
also, that lattice results suggest that this is approximately
true at small $N_c$.
The appearance of eigenvalue repulsion is clear when the spatial
volume is small, as when the theory lives on a very small sphere.
At infinite $N_c$,
Aharony {\it et. al.}\cite{aharony} showed that the effective theory
only involves a constant mode, and reduces to a 
random matrix model for that mode.
As typical of random matrix models, eigenvalue repulsion 
arises from the Vandermonde determinant in the measure of the group integral.

In infinite volume, instead of one random matrix,
there is a field theory of (not quite) random matrices.  Such field theories
have been studied little.  The
interesting, and gauge invariant, question is how
confinement arises from the {\it dynamical} generation of eigenvalue
repulsion.

\section{Acknowledgements} 

This research was supported by D.O.E.
grant DE-AC02-98CH10886, and in part by
the Alexander von Humboldt Foundation.  I thank Prof. T. Kunihiro,
and the other organizers of YKIS 2006, for their invitation and
warm hospitality.

\end{document}